\documentclass[11pt,twoside]{article}
\usepackage{./asp2014}
\usepackage{color}

\aspSuppressVolSlug
\resetcounters

\begin{document}

\title{Current State of Astrophysical Opacities: A White Paper}
\author{A.~E.~Lynas-Gray,$^{1, 2}$ S.~Basu,$^3$ M.~A.~Bautista,$^4$ J.~Colgan,$^5$ C.~Mendoza,$^4$ J.~Tennyson,$^1$ R.~Trampedach,$^{6, 7}$ and S.~Turck-Chi{\`e}ze$^8$ \\
\affil{$^1$Dept. of Physics and Astronomy, University College London, London WC1E 6BT, UK}
\affil{$^2$Dept. of Physics, University of Oxford, Oxford OX1 3RH, UK; \email{tony.lynas-gray@physics.ox.ac.uk}}
\affil{$^3$Dept. of Astronomy, Yale University, CT 06511, USA}
\affil{$^4$Dept. of Physics, Western Michigan University, Kalamazoo, MI 49008, USA}
\affil{$^5$Los Alamos National Laboratory, Los Alamos, NM 87545, USA}
\affil{$^6$Space Science Institute, Boulder, CO 80301, USA}
\affil{$^7$Stellar Astrophysics Centre, Aarhus University, 8000 Aarhus, Denmark}
\affil{$^8$CEA/IRFU/SAp, CE Saclay, F-91191 Gif sur Yvette, France}
}

\paperauthor{A.~E.~Lynas-Gray}{tony.lynas-gray@physics.ox.ac.u}{}{University College London}{Dept. of Physics and Astronomy}{London}{}{WC1E 6BT}{UK}
\paperauthor{Sarbani~Basu}{sarbani.basu@yale.edu}{0000-0002-6163-3472}{Yale University}{Department of Astronomy}{New Haven}{CT}{06520-8101}{USA}
\paperauthor{Manuel~Bautista}{manuel.bautista@wmich.edu}{}{Western Michigan University}{Dept. of Physics}{Kalamazoo}{MI}{49008}{USA}
\paperauthor{Claudio~Mendoza}{claudio.mendozaguardia@wmich.edu}{}{Western Michigan University}{Dept. of Physics}{Kalamazoo}{MI}{49008}{USA}
\paperauthor{J.~Colgan}{jcolgan@lanl.gov}{}{Los Alamos National Laboratory}{}{Los Alamos}{NM}{87545}{USA}
\paperauthor{Jonathan~Tennyson}{j.tennyson@ucl.ac.uk}{0000-0002-4994-5238}{University College London}{Department of Physics and Astronomy}{London}{}{WC1E 6BT}{UK}
\paperauthor{Regner Trampedach}{rtrampedach@SpaceScience.org}{}{Space Science Institute}{}{Boulder}{CO}{80301}{USA}
\paperauthor{Sylvaine~Turck-Chi\`eze}{sylvaine.turck-chieze@cea.fr}{}{DAP/IRFU/DRF/CEA, UMR AIM}{Universit\'e Paris-Saclay, CE Saclay}{Gif sur Yvette}{}{91191}{France}

%\aindex{Turck-Chi\`eze, S.}
%\aindex{Trampedach, R.}
%\aindex{Tennyson, J.}
%\aindex{Basu, S.}
%\aindex{Lynas-Gray, A.~E.}
%\aindex{Bautista, M.}
%\aindex{Mendoza, C.}
%\aindex{Colgan, J.}

\begin{abstract}
  Availability of reliable atomic and molecular opacity tables is essential in a wide variety of astronomical modeling: the solar and stellar interiors, stellar and planetary atmospheres, stellar evolution, pulsating stars, and protoplanetary disks, to name a few. With the advancement of powerful research techniques---such as helioseismology and asteroseismology, solar neutrino-flux measurements, exoplanet survey satellites, three-dimensional hydrodynamic atmospheric simulations (including non-LTE and granulation effects), high-performance computing of atomic and molecular data, and innovative plasma experiments---the accuracy and completeness of opacity tables is being taken to an unprecedented level.  The goal of the second Workshop on Astrophysical Opacities was to gather opacity data producers and consumers from both the atomic and molecular sectors to contribute to solving outstanding problems and to develop more effective and integrated interfaces. In this review we attempt to summarize the discussion at the workshop and propose future directions for opacity research.
\end{abstract}

%++++++++++++++++++++++++++++++++++++++++++++++++++++++++++++++++++++++++++++++
\section{Introduction}
%++++++++++++++++++++++++++++++++++++++++++++++++++++++++++++++++++++++++++++++

The study, analysis, and generation of astrophysical opacities \citep[e.g.,][Turck-Chi\`eze, this volume and references therein]{edd32, eps51, sea59, cox65a, cox65b, 1987JPhB...20.6363S, 1987JPhB...20.6379B, 1990SoPh..128...49C, rog92a, rog92b, 1993ApJ...408..347T, rog94, 1994MNRAS.266..805S, igl95, 1996ApJ...464..943I, 1998SSRv...85..125T, 2005MNRAS.360..458B, 2007ApJS..168..140S, 2012MNRAS.425...21T, mon15, 2015Natur.517...56B, 2016MNRAS.458.1427H, 2016ApJ...817..116C, 2016ApJ...821...45K, 2016ApJ...824...98K, pain:DetailOpacCalcs} is a complex and demanding pursuit. So much so that only a handful of groups around the world have the stamina, incentive, expertise, and human and computational resources to actively work in this field. Computing opacities \citep{2017arXiv170403528M} involves knowledge of atomic and molecular physics in order to calculate energy levels, line positions, $f$-values, and photoabsorption cross sections for large numbers of ions, atoms, and molecules of astrophysical interest; computational expertise to perform massive computations and to process large data volumes; and proficiency in plasma physics for the development of equations of state to determine ionization fractions and energy level occupation numbers, as well as line-broadening processes, all under extreme temperature and density regimes.

Validating the quality and accuracy of opacity tables is complicated by the fact that the opacity itself is not directly observable, and can only be inferred from radiative transfer calculations (although simplified limiting cases often apply).
Moreover, stellar interior conditions are difficult to achieve in a laboratory and for a long time were limited to external layers of massive stars (see a recent comparison of calculations for this range of conditions by  \citealt{2016ApJ...823...78T}). Despite a long history of opacity experiments \citep[see, for instance,][for a review
of experiments at conditions relevant for the solar convective envelope]{pain:DetailOpacCalcs}, plasma conditions approaching those of the solar radiative zone have only been recently reached \citep{bailey:Fe150eVplus,2015Natur.517...56B}. Apart from the very limited number of experimental opacities available at present, it is only through applications in the study of the solar
interior, stellar pulsations, and the atmospheric spectra of stars and planets that theoretical opacities can be more or less directly confronted with reality.

Nonetheless, present needs for improvements in current theoretical opacity data can hardly be overstated. First, we know that the combination of current opacities with the revised solar photospheric abundances of \citet[][AGSS09]{2009ARA&A..47..481A} fail to give standard solar models within the constraints of helioseismic observations \citep[][and Section~4 below]{2004PhRvL..93u1102T,
serenelli:SunSeismAGSS09,2010ApJ...715.1539T}.  Current studies of mode instabilities in pulsating stars seem to require opacities significantly larger than those in current opacity tables \citep{das10}. Modern studies of cool planetary atmospheres, on the one hand, and hot accretion disks around compact objects and protostars, on the other, require opacities well outside the range of most current tables (e.g., of dust and aerosols). Furthermore, the recent discovery of a neutron-star merger  \citep{2017ApJ...850L..40A}, which brings about rapid neutron-capture nucleosynthesis, has shown the need for heavy-element opacities that are simply unavailable.

The Workshop on Astrophysical Opacities (WAO) provides a forum for producers and consumers of opacity data from both the atomic and molecular sectors to contribute to solving outstanding problems and to develop more effective and integrated interfaces. The previous WAO took place at the IBM Venezuela Scientific Center, Caracas, Venezuela, in July 1991 \citep{1992RMxAA..23.....L}. This was an iconic event in the field as it brought together the two main groups recomputing the atomic astrophysical opacities at the time, the OPAL group \citep{igl92a, igl92b, rog92} and the Opacity Project \citep{sea92}, and it was at or around that meeting that complete opacity tables were first released by both teams.

There have been major advances in astrophysical opacities since that first WAO, to which the subsequent WorkOp workshop series \citep{2000JQSRT..65..527S} has contributed. The strict constraints now imposed on atomic opacities by the helioseismic and solar neutrino benchmarks \citep[and references therein]{tur16} have led to a proliferation of opacity codes, new opacity tables \citep[e.g., OPLIB,][]{2016ApJ...817..116C}, and extensive comparisons \citep[for the iron-group bump, see][]{gil12, gil13, tur13} to examine key issues (completeness, configuration interaction, line broadening, plasma effects) and to look for possible sources of ``missing'' opacity. On the other hand, direct and transit spectroscopic observations of hundreds of exoplanets detected by space probes such as  {\it Kepler}\footnote{\url{http://kepler.nasa.gov/}} and {\it Corot}\footnote{\url{http://bit.ly/2cYJ09w}} have uncovered a peremptory demand for molecular opacities \citep{ber14}.

Along with these developments we have also seen the rise of asteroseismology as a window on stellar interiors \citep{jcd:AsteroseismReview}, large-scale surveys like the Two Micron All Sky Survey \citep[2MASS,][]{skrutskie:2MASS}
and the Sloan Digital Sky Survey \citep[SDSS,][]{blanton:SDSS-IV}, and the astrometry mission Hipparcos  \citep{vanleeuwen:HipparcosNewReduc} currently being extended to a billion stars in the Galaxy and beyond by the GAIA mission
\citep{lindegren:GAIA-DR1}, all providing much tighter constraints on our modeling efforts. Thus, we were encouraged to organize a second WAO that took place at Western Michigan University, Kalamazoo, Michigan, USA, during the week 1--4 August 2017.

In practice the opacities needed in stellar astrophysics must span a wide range of conditions from a star's surface to its center. In terms of temperature and density, conditions range from $10^3$~K and $10^{-12}$~g\,cm$^{-3}$ to  $10^9$~K and $10^9$~g\,cm$^{-3}$. The more recent needs of planetary atmosphere research extend the temperature range down to a few hundred Kelvin, where molecules dominate over atomic species. In modeling stellar interiors it is safe to assume local thermodynamic equilibrium (LTE), which allows us to compute opacities from collisional and radiative data for all processes involved in energy transport, while ionization equilibria and level populations are set by an assumed equation of state (EOS). However, two big complications arise. (1) The enormous amount of atomic (and ionic) data needed to account for all bound levels from the ground to highly excited ones, and to include all possible transitions between them and the continuum, which brings about possible issues of completeness and propagation of uncertainties in the radiative data. Towards the lower temperatures of planetary atmospheres, the emphasis shifts from atoms to molecules expanding the size of cross-section data volumes by orders of magnitude. (2) The adoption of an EOS valid for the same large range of plasma conditions that includes all relevant processes, in particular high-density plasma interactions (see Section 3), and propagating the results to the opacity calculation for a consistent treatment. In the lower density regime, as in accretion disks and some stellar atmospheres, the LTE approximation breaks down and collisional cross sections are needed, as well as the photoabsorption data for individual transitions.

The present report attempts to summarize the discussions during the second WAO and suggest future priorities for opacity research.  Some bias in the views expressed is inevitable given work already undertaken by the authors to compute atomic and molecular data, as well as opacity tables. Readers need to be aware of the Lawrence Livermore National Laboratory (LLNL) work not fully represented in the present paper, and the perspective on future directions for opacity research that follow. But all who are engaged in calculating opacity tables share a common objective: to provide opacity tables of the highest achievable accuracy for present and future astrophysics research.

%++++++++++++++++++++++++++++++++++++++++++++++++++++++++++++++++++++++++++++++
\section{Workshop on Astrophysical Opacities}
%++++++++++++++++++++++++++++++++++++++++++++++++++++++++++++++++++++++++++++++

One of the main highlights of the second WAO was the discussion on the competing needs for accuracy and completeness. Completeness gives priority to identifying and taking account of all relevant physical processes that contribute to the opacity in the astrophysical context. Accuracy insists on minimizing assumptions and approximations to model already known physical processes as reliably as possible. Of course it was recognized that both are needed, but resource limitations demand priorities be assigned. The debate was motivated in great part by the realization that astrophysical applications, such as modeling the solar structure and stellar pulsation, and recent laboratory opacity determinations all seem to suggest that current theoretical opacities may be significantly underestimated. Hence, the debate centered around whether focusing on completeness would provide improved opacities more rapidly (and cheaply) than expending effort on accuracy.

The many excellent papers reporting on the WAO in the present proceedings is a testimony to the current scope, sophistication, and detail of opacity calculations. Also their applications and verifications through astrophysical modeling and measurements under a range of relevant plasma conditions. We refer the reader to these papers for full details. Instead of summarizing each
contribution, we review opacities currently in use and give reasons for an essential validation.

\citeauthor{1982ApJ...260L..87S}'s \citeyearpar{1982ApJ...260L..87S} plea for a reexamination of metal opacities in stars led to the Opacity Project \citep[OP,][]{1994MNRAS.266..805S,2005MNRAS.360..458B} and OPAL \citep{1996ApJ...464..943I} revisions to the astrophysical opacities then in use. A high priority for both projects were physically consistent and much improved EOS foundations for the opacity calculations, developed independently and in two conceptually different formulations: the OPAL EOS by \citet{2002ApJ...576.1064R} in the \emph{physical picture} and the MHD EOS by \citet{1988ApJ...331..794H}, \citet{1988ApJ...331..815M}, \citet{1988ApJ...332..261D}, and \citet{1990ApJ...350..300M} in the \emph{chemical picture}. Efforts were made to identify and take account of all atomic and plasma processes of importance to the calculation of stellar envelope opacities. They were nonetheless limited by the computing power at the time and approximations were made, many of which can now be relaxed.

Astronomical evidence for the need to further revise stellar opacities comes from several lines of inquiry: the \emph{missing opacity in the Sun problem} labels the issue that solar models based on current atomic opacities combined with the low-metal abundances of AGSS09 cannot be made to agree with the helioseismic constraints \citep{serenelli:SunSeismAGSS09}. Agreement with helioseismology could be restored by either increasing the atomic opacity by some 20\% at the bottom of the convection zone and decreasing it towards the solar center \citep{serenelli:SunSeismAGSS09, jcd:AsteroSeism} or by returning to the older, higher metal abundances. The AGSS09 abundances cannot, however, be easily dismissed. They constitute the first homogeneous abundance analysis of the Sun (as opposed to a compilation of a plethora of other groups' works); they are based on a 3D solar atmosphere simulation that greatly outperforms the semi-empirical Holweger--M{\"u}ller model in reproducing solar observations \citep{tiago:suneHlinesLimbdark}; and the 3D atmosphere gives asymmetric line profiles that match observations and reveal even weak blends that go unnoticed in a 1D analysis \citep{grevesse:StopUsingGN93}. Moreover, the accuracy of seismic results has highlighted the need to verify opacity calculations in an attempt to reconcile the Standard Solar Model, helioseismology, and solar neutrino fluxes (see the review by \citealt{2011RPPh...74h6901T}).

On the other hand, the AGSS09 abundances are disputed by \citet{caffau:CLSFB11-SunAbunds}, who find abundances much closer to the classic values, based on an independent 3D solar atmosphere simulation but analyzing only five of the 15 metals included in the \citet[][OP05]{2005MNRAS.360..458B} opacities. Higher metal abundances have also been recently derived from {\it in situ} measurements of the solar wind by \citet{ste16}, but in spite of an improved accord with the helioseismic sound speed around the base of the convection zone, they have been questioned for leading to neutrino overproduction due to an excess of refractory elements, namely, Mg, Si, S, and Fe \citep{ser16, vag17a, vag17b}. The connection between photospheric and interior compositions is also debatable.

\citet{2012MNRAS.422.3460S} have tested the dependence of the Magellanic Cloud B~star pulsation excitation on opacity and chemical mixture. With known Magellanic Cloud abundances, they find these pulsations can be driven by an increased iron-group bump opacity as long as this occurs at the temperature corresponding to the maximum nickel contribution. An additional astronomical hint comes from a study of $\beta$\,Cephei pulsation by \citet{2017EPJWC.15206005W} and  \citet{2017MNRAS.466.2284D}, who show for $\nu$\,Eri, $\gamma$\,Peg, and 12\,Lac how a modified opacity would improve the agreement between observed and calculated pulsation frequencies. This includes a ${\sim}50$\% opacity increase around $T=290$\,kK, which is a larger change at a lower temperature than that needed to resolve the missing opacity in the Sun.

A further indication that current atomic opacities may need an increase comes from laboratory measurements by \citet{2015Natur.517...56B}, who used the Sandia Z-Facility to create an iron plasma at conditions close to those at the base of the solar convection zone and measured its opacity in transmission.  The iron absorption measured by \citeauthor{2015Natur.517...56B}\ is demonstrably higher and broadened when compared with current theoretical predictions.   For a solar composition at a temperature of $T = 2.11 \times 10^6$~K and electron density of $n_e = 3.1 \times 10^{22}$~cm$^{-3}$,
\citeauthor{2015Natur.517...56B} replaced the iron contribution to the OP opacity with their experimental determination (in the wavelength range $7 < \lambda < 12$~{\AA}), and showed that this alone increased the Rosseland mean opacity by $7\pm 3$\%.
If correct, the higher opacities implied by the \citeauthor{2015Natur.517...56B}\ experiment strongly indicate that we are missing physics in current opacity models. From the theoretical view point, though, it seems unlikely that more accurate models or larger calculations of the same type will lead to an agreement with this measurement.  The experimental data appear to contravene the oscillator strength sum rule \citep{2015HEDP...15....4I}, which implies that an overlooked mechanism might be responsible for the result. For example, \cite{mor17} have proposed two-photon contributions as a source of the missing opacity; however, this possibility is very uncertain \citep{2018HEDP...26...23P}, and it is also unclear why previous comparisons of opacity calculations and measurements did not require additions to existing models to concur. The \citeauthor{2015Natur.517...56B}\ experiment needs to be replicated using a different setup; \citet{2017JPlPh..83a5903H} discuss the prospects for doing so using the National Ignition Facility (NIF).

\citet{2016PhRvL.116w5003N} carried out an extensive close-coupling calculation of \ion{Fe}{xvii}, a major contributor to the opacity at the base of the solar convection zone and the dominant ion present in \citeauthor{2015Natur.517...56B}'s experiment. \citeauthor{2016PhRvL.116w5003N} demonstrate, as had already been done previously \citep{1995ApJ...443..460I}, that the much simpler calculations performed under the original OP \citep{1994MNRAS.266..805S} severely underestimated this absorption. However, whether such close-coupling calculations can help resolve current discrepancies between the solar model and the experiment by \citeauthor{2015Natur.517...56B}\ remains an intensely debated topic \citep{2016PhRvL.117x9501B}. As atomic
core excitations and ionizations are accounted for using Voigt profiles---or in some cases a far-wing cut-off following \citet{1994MNRAS.266..805S} in other current opacity calculations \citep{2017ApJ...835..284I} including the more recent OP opacities by \citet{2005MNRAS.360..458B}---it is not clear to what extent the more elaborate, but so far less complete cross sections by \citeauthor{2016PhRvL.116w5003N}, will increase the Rosseland mean of a solar mixture.

The more physical close-coupling formulation results in Fano-shaped resonances effectively moving absorption from  the line cores to the wide wings, increasing the harmonic Rosseland mean despite identical total oscillator strengths. Collisional line broadening effects can often hide subtle Fano line-profile effects at moderate and high plasma densities. But whether absorption is moved to, or away from, windows in the absorption by other elements of a solar mix determines whether the Rosseland mean will increase and possibly resolve the disagreement between calculations and experiment \citep{2015Natur.517...56B}. Moreover, it remains to be demonstrated if close-coupling calculations alone, however accurate, can achieve the required completeness for the more complicated ions. Conversely, perturbative approaches can account for many more transitions, but some caution is advised in that they may miss important coupling effects and introduce systematic errors in the mean opacities. Adopting a hybrid model may provide a solution as configuration interaction only affects some transitions.

%++++++++++++++++++++++++++++++++++++++++++++++++++++++++++++++++++++++++++++++
\section{Plasma Effects and Line Broadening}
%++++++++++++++++++++++++++++++++++++++++++++++++++++++++++++++++++++++++++++++

Plasma effects refer to phenomena arising from collisions of finite duration between particles in a plasma rendering the gas non-ideal. (Thermodynamics rely on collisions for thermalization but are assumed infinitely short in ideal gases, and all plasma effects mentioned below are therefore incompatible with the ideal gas assumption.) The most well-known effect of non-ideal collisions is line broadening, either by Stark broadening of resonance lines or by perturbing bound states on time scales shorter than the radiative lifetime that increase their uncertainty in energy and, hence, their line widths.

A bound state may be so perturbed by collisions that the electron can no longer be assumed bound, and has effectively been removed from the parent ion causing a second plasma effect known as {\it pressure ionization}. \citet{1988ApJ...331..794H} introduced the occupation probability formalism to describe this process in an EOS based on the chemical picture by minimizing the Helmholtz free energy, which also makes partition functions converge. In the physical picture the activities of clusters of electrons and nuclei consist of bound and scattering state parts that together are convergent. No assertions about occupation probabilities need to be made, and non-Boltzmann level populations arise directly from the formulation. Regardless of the EOS framework employed, this effectively moves absorption from lines to the continuum (bound--bound to bound--free) as states pressure ionize  \citep{1987ApJ...319..195D}. Such heavily perturbed states are often referred to as \emph{dissolved states}.

A third plasma effect arises when electronic wave functions are perturbed resulting in shifts in energy levels and line positions, energy-level shifts being larger than line shifts since the former largely cancel in differencing. There have been persistent measurements of such shifts, most recently by \citet{hansen:DenseFeEshifts} for iron. Energy-level shifts (continuum lowering) are, however, much less than expected \citep{1988ApJ...331..794H} if that alone were responsible for bound states dissolving into the continuum \citep{rogers:BndStatesDebyePot}.

A fourth plasma effect is that on the free energy, to first order described by \citet{Deb:Huck} theory, due to the screened Coulomb interactions. In the physical picture the Debye--H{\"u}ckel term is the sum of all ring diagrams in the activity expansion. This was the single most important stellar EOS improvement at the time, which brought solar models into agreement with a helioseismic analysis of the convection zone \citep[where opacities have no  effect,][]{jcd-wd:osc-eos}.

All four plasma effects must be operating simultaneously and connected by the probability distribution of the electric field strength from passing ions, the so-called {\it microfield distribution function} (the screened Coulomb coupling affects the microfield distribution). There are several choices of microfield distributions in the literature \citep[e.g.,][]{1999ApJ...526..451N, iglesias:MultiCompMicrofield, potekhin:microF, laulan:FastQbeta}, but little observational or experimental guidance as to which is preferred. Occupation probabilities also depend on the adopted critical field strength for ionization, which depends on the exact mechanism considered \citep{luc-koenig:StarkContinuum, 1988ApJ...331..794H}, adding an uncertainty that is convolved with the microfield distribution. The level shifts and occupation probabilities are usually considered mutually exclusive, and a consistent picture of the combined effects is needed. Considerable progress has been made towards this goal through the ChemEOS model \citep{chemeos1,chemeos2}, which is implemented in the ATOMIC code and used to make the recent OPLIB tables \citep{2016ApJ...817..116C}. ChemEOS minimizes the free energy in a chemical picture, and makes use of an occupation probability formalism that is generated from a microfield distribution \citep{potekhin:microF}. ChemEOS also goes beyond the Debye--H{\"u}ckel approximation in the Coulomb contribution to the free energy.

Hydrogen lines in white-dwarf atmospheres provide strong observational tests of how lines dissolve into the continuum. For that purpose \citet{2009ApJ...696.1755T} analyzed Balmer spectra of 250 DA white dwarfs with non-convective (and therefore simpler) atmospheres. By employing occupation probabilities, they found greatly improved consistency among Balmer lines and masses in better agreement with stellar evolution. They also found that a more realistic implementation of the microfield distribution function \citep{1999ApJ...526..451N} is needed, rather than the approximation to the Holtsmark distribution originally used by \citet{1988ApJ...331..794H}. However, there is still some disagreement between white-dwarf masses determined from such spectra {\it vs.}~from photometry or gravitational redshifts.

The asymmetric structure near the core of Stark-broadened lines is affected by the order of the multipole expansion used for describing the perturbing charges. \citet{2016PhRvA..94b2501G} investigated its convergence, and found profiles diverge from the dipole approximation with increasing perturber density (both cores and wings are affected). Line cores are more important
for interpreting experiments or stellar spectra, but the wings will also affect opacity calculations in general. With a conserved total oscillator strength (from sum rules), redistributing absorption from line cores to the wings is an efficient way of increasing the harmonic mean of the Rosseland opacity.  Experiments by \citet{2016PhRvA..94b2501G} and \citet{2017ASPC..509..149F} provide strong experimental constraints on our models, and efforts are underway to derive experimental occupation probabilities.

\citet{2016JPhCS.717a2069M} notes that spectral line shapes in large-scale opacity calculations involving L- and M-shell transitions are modeled as Voigt profiles, whereby detailed ion broadening effects are ignored. But an ion's microfield distribution can lead to forbidden (field-dependent) line transitions and significant changes to the broadening and shape of line profiles. \citeauthor{2016JPhCS.717a2069M} presents Stark broadening calculations for line profiles of L-shell transitions, linking the ground state and singly excited states in Ne-like iron ions, at the plasma conditions \citet{2015Natur.517...56B} adopted for their Z-Facility experiment. It is clear that forbidden components and altered line profile shapes will affect the opacity through an absorption frequency redistribution.

If not modified, the far red line wings of Lorentz-shaped resonance lines will dominate the opacity at low temperatures. \citet{1994MNRAS.266..805S} employed an ad-hoc $\lambda^{-4}$ suppression, but the line-wing satellites from
H--H$^+$ collisions \citep{allard:quasi-H2-Lya} provide a more physical and exponential suppression.

In doubly excited atoms/ions, excited spectator electrons can cause interference with an electron transition forming a line \citep{2010HEDP....6..318I} thus altering its line shape. This process is crucially affected by the microfield distribution suppressing the Rydberg series of spectator electrons. Therefore, the Fano profiles of autoionization lines must also be subject to broadening, but the theoretical basis for this is only now being established (Pradhan \& Nahar, this volume).

The many kinds of transitions giving rise to spectral lines, combined with the many broadening agents, lead to a large variety of line shapes. For this reason simplified profiles will be inadequate in many cases, although we also need tractable calculations in light of the (at least) tens of millions of lines involved in opacity calculations. A key ingredient is therefore fast and efficient yet accurate algorithms \citep{sonnad:FastLineCalc}.

It is clear that there are already many efforts to address these issues from stellar observations, ambitious laboratory experiments, and sophisticated quantum mechanical calculations. There is still much to be learned, and these efforts must therefore be encouraged.

%++++++++++++++++++++++++++++++++++++++++++++++++++++++++++++++++++++++++++++++
\section{Missing Opacity in the Sun}
%++++++++++++++++++++++++++++++++++++++++++++++++++++++++++++++++++++++++++++++

The issue of whether our current opacity calculations are accurate is critical for solving a major problem in our solar models: Why is the structure of solar models constructed with the latest solar abundances of AGSS09 so different from that of the Sun? The structure mismatch is most notable at the base of the convection zone, and results imply that the opacity at such conditions must be higher than that included in the models.

Opacity has of course two components: first the intrinsic opacity at the density and temperature of a given chemical composition, and second, the heavy element composition. For a given intrinsic opacity, the total opacity increases with metallicity. Helioseismic data, which probe the structure of the Sun, are only sensitive to the sound speed and density in a given region and, hence, cannot distinguish between total and intrinsic opacities (see Basu, this volume). Models constructed with older, higher metallicities \citep{1998SSRv...85..161G} agree very well with the Sun. This has led to what
is often described as the ``missing opacity in the Sun problem,'' but it is most likely to be a problem with total opacities. Sorting out whether the issue is with intrinsic opacities or metallicity is critical because solar metallicities are used as the standard for describing the metal content of other stars. Metallicity affects both the structure and evolution of a star: low-metallicity models evolve faster. Thus errors in solar metallicities will result in errors in the ages of astrophysical objects such as exoplanet hosts or stellar clusters.

Increasing intrinsic opacities as tabulated by OPAL, OP, etc. could solve the problem with the structure mismatch of the solar models and the Sun. However, the increment must be for elements that contribute the most at the solar convection-zone base (O, Fe, Ne, and Si in that order), and it should taper off towards the solar core; an overall increase would merely move the problem to the solar core. Thus clearly there is more work to be done. Experiments by \citet{2015Natur.517...56B} are important pieces of the puzzle, and interpreting those results in terms of opacity calculations is central to resolving this and the other issues noted above.

We currently know of two ways to test interior abundances independently of our knowledge of opacities. Firstly N, O, and F in the core produce neutrinos (as part of the CNO cycles) that have energy thresholds between those of the pp and B neutrinos \citep[see, for instance,][]{bahcall:newOP+AGS05}, which means their fluxes can be measured separately. However, current neutrino measurements have so far only produced upper limits to their fluxes \citep{bergstroem:SolarNeutrFlux2016}.
Secondly the ionization effect on the adiabatic exponent, $\gamma_1$, leaves little seismic glitches in the convection zone where stratification is determined by $\gamma_1$. \citet{lin:seismEOS} found that groups of elements with similar ionization energies make bumps in $\gamma_1$ that can be marginally constrained with helioseismology, but individual metals are not
accessible. \citet{vorontsov:SeismicEOScalibr} found support for generally lower metallicity from a similar helioseismic analysis.

Despite much progress, state-of-the-art solar models \citep{2017ApJ...835..202V} still leave out processes that might be significant regarding the missing opacity in the Sun problem, most notably mixing mechanisms that can operate in the (nominally) convectively stable regions. This includes overshooting from convective regions \citep{rempel:OvershootModel} into Schwarzschild-stable regions; semi- or double-diffusive convection \citep{wood:SemiConvMixing}; internal gravity waves \citep{talon:Int-g-waves1}; the combined effects of gravitational settling; thermal diffusion and radiative levitation \citep{turcotte:Diff+RadLev}; and rotational mixing \citep{eggenberger:RotMixSunType}. Furthermore, stellar models do not in general include a non-grey stellar atmosphere \citep{trampedach:T-tau}. Many improvements can therefore still be applied to our solar modeling, but none of these is likely to solve the missing opacity in the Sun problem---not even in aggregate. Thus improving, or at least validating, opacity calculations and measurements will remain the key to solving the problem. Comparisons between recent and earlier opacity calculations show that element completeness leads to an increase of about 20\% for iron and nickel but a decrease for other elements. OPAS opacity tables \citep{2015ApJ...813L..42L}, for example, give an increased mean opacity at the base of the convection zone of 6\%, about a third of what is required to remove the disagreement with helioseismology.

%++++++++++++++++++++++++++++++++++++++++++++++++++++++++++++++++++++++++++++++
\section{Molecular Opacities}
%++++++++++++++++++++++++++++++++++++++++++++++++++++++++++++++++++++++++++++++

Molecules provide the dominant source of opacity in cool stars and brown dwarfs. The complexity of molecular spectra often leads to whole-scale blanketing of large spectral regions, which can have a profound effect on the properties of the atmosphere \citep{84ErGuJo.HCN}. Molecular opacities are dominated by bound--bound transitions, so are normally based on extensive line
lists of molecular transitions. The construction of such line lists was in its infancy at the time of the Caracas WAO \citep{92Jorg}, but this situation has now changed radically (see papers by Bernath, Calvet, Huang et al., Morley, Sousa-Silva et al., Tennyson, and Zammit et al. in this volume). There are coordinated projects such as ExoMol \citep{2012MNRAS.425...21T} and TheoReTS \citep{TheoReTS} leading to systematic compilations of molecular line lists for modeling hot atmospheres \citep{HITEMP, TheoReTS, 2016JMoSp.327...73T}. Much of this work has been stimulated by the desire to model and characterize exoplanets. As for atomic opacities, issues to do with completeness against accuracy are important, with models showing that for key species (e.g., methane) correct results can only be obtained by considering many billions of transitions \citep{jt572}.

Although there has been a lot of work constructing extensive and reliable molecular line lists and hence opacities, there remains much to be done. BT-Settl model atmospheres \citep{BT-Settl} are among the most used for cool stars and brown dwarfs. In their study of BT-Settl model atmospheres for M-dwarfs, \citet{13RaReAl.NaHAlH} identified only AlH, NaH, and CaOH as key
species for which data were missing. The ExoMol project has since provided line lists for AlH \citep{jtAlH} and NaH \citep{jt605}, but CaOH remains outstanding. The chemistry of carbon stars is more complicated than oxygen rich ones, and data for a number of species are required to complete good models, foremost amongst these are acetylene (HCCH) and C$_3$.

Transition metal containing diatomics such as TiO provide strong sources of opacities since these open-shell species often display strong electronic transitions in the near infrared and visible, i.e., near the stellar flux peak for cool stars. There are some line lists, of variable accuracy, available for key species such as  TiO, VO, FeH, ScH, and TiH, but data are missing for many others such as CrH, NiH, ZrO, YO, and FeO, to name a few. Constructing  accurate theoretical line lists for transition metal containing molecules remains very challenging \citep{jt623,jt632}.

As mentioned, the field of exoplanets has been a major driver for extending studies of molecular opacities. \citet{1602.06305} recently published a review of laboratory data needs for work on exoplanet atmospheres that covers opacities and much else.  In particular, the recent discovery of hot rocky exoplanets, also known as super-Earths and lava planets, has raised the need to study a whole range of new molecular species. The possible molecules important in the atmospheres of these bodies, which can reach temperatures over $3000\,{\rm K}$, have been reviewed by \citet{jt693}.

%++++++++++++++++++++++++++++++++++++++++++++++++++++++++++++++++++++++++++++++
\section{Opacity Due to Trans-Iron Elements}
%++++++++++++++++++++++++++++++++++++++++++++++++++++++++++++++++++++++++++++++

Scarcely a fortnight after the Kalamazoo WAO, the Laser Interferometer Gravitational-Wave Observatory (LIGO) detected gravitational waves from the binary neutron-star (NS) merger GW170817 \citep[LIGO Scientific Collaboration\footnote{\url{https://www.ligo.org/partners/}}
\& Virgo Collaboration\footnote{\url{http://public.virgo-gw.eu/the-virgo-collaboration/}}] {2017ApJ...850L..39A,2017ApJ...850L..40A}. This discovery was jointly observed in the UV, optical, and IR with various ground- and space-based instruments \citep{2017ApJ...848L..19C, 2017ApJ...848L..17C, 2017ApJ...848L..27T} opening a new era in astronomy in which atomic opacities are again at the forefront of cutting-edge science.

Electromagnetic signatures accompanying NS mergers, referred to as macronovae \citep{2005astro.ph.10256K} or kilonovae \citep{2010MNRAS.406.2650M}, are roughly isotropic thermal transients powered by radioactive decay of r-process
elements synthesized in the ejecta \citep{1998ApJ...507L..59L, 2011ApJ...736L..21R, 2012ApJ...746...48M, 2013ApJ...775..113T}. Furthermore, NS mergers are now believed to be the main site for rapid neutron-capture nucleosynthesis of heavy elements in the universe. In such NS merger events the time scale and shape of the transient light curve is determined by the fireball opacity.
It is believed that lanthanides and, additionally, actinides can enhance the opacity by nearly two orders of magnitude over that of typical iron-rich supernova ejecta \citep{2013ApJ...774...25K,2013ApJ...775..113T}, resulting in a hotter, denser, and longer lived fireball.

While a start has been made \citep{2015HEDP...16...53F, fon17} in providing opacity tables to characterize this type of heavy-element rich expanding ejecta, most kilonova models assume grey (wavelength independent) opacities. In modeling the light curve of GW170817, \citet{2017ApJ...848L..17C} tested opacity values between $\kappa = 0.1$~cm$^2$\,g$^{-1}$ (appropriate for Fe-peak ejecta) and $\kappa = 10$~cm$^{2}$\,g$^{-1}$ \citep[estimated for lanthanide-rich ejecta by][]{2013ApJ...774...25K}. They found that at least two independent components had to be used to roughly reproduce the UV/optical/IR light curves, and these would need an opacity of $\kappa\simeq 3$~cm$^{2}$\,g$^{-1}$.

Many more events like GW170817 will be observed in the near future. These observations may allow us to answer a number of fundamental questions in the physics and astronomy pertaining to cosmology, chemical evolution, nuclear physics, and fundamental atomic physics, to name just a few. Study of GW170817 and other anticipated NS merger observations points out the compelling need to compute reliable opacities for heavy-element rich plasmas. Many of the methods and codes developed for stellar opacities over the last few decades (for example, TOPAZ, \citealp{2003JQSRT..81..227I}) can become the basis for such work, but additional theoretical and numerical methods are needed in some cases to deal with the fully relativistic and very complex atomic structure of such heavy elements.

Heavy elements can have several bound electrons in excited states, all with non-negligible abundances due to a large configuration degeneracy not significantly reduced by Boltzmann factors. The {\it super-transition array} (STA) formalism
\citep[and references therein]{1995PhRvE..52.6686B,1997PhRvE..56...70B} was developed to address the challenge presented by several bound electrons in excited states.  While the STA method is used by several research groups, hybrid calculations combining detailed line accounting with STA are also being performed \citep[and references therein]{2015HEDP...15...30P, 2007HEDP....3..109H}. \citet{igl95} and \citet{2016ApJ...821...45K} have studied the impact of heavy elements on stellar opacities using the STA formalism, and find that, while these elements are effective photon absorbers, their low abundances lead to an insignificant contribution to the solar interior Rosseland mean opacities.

Finally, methods developed for astrophysical opacities can also be used when atomic data are required for other applications.  For example, the atomic data requirements for accurate modeling of tungsten plasmas that may contaminate current
and planned magnetic fusion devices are formidable \citep{tungsten1}, and the problem is also greatly complicated by the non-LTE nature of fusion plasmas \citep{tungsten2}. Also accurate atomic data are required to model tin plasmas generated for EUV lithography applications \citep{2017HEDP...23..133C}. Depending on how these laser-produced plasmas are generated, resulting plasmas may or may not be in LTE.

%++++++++++++++++++++++++++++++++++++++++++++++++++++++++++++++++++++++++++++++
\section{General Findings}
%++++++++++++++++++++++++++++++++++++++++++++++++++++++++++++++++++++++++++++++

With regard to the future improvement of astrophysical opacities, in our opinion efforts need to take into account the following considerations:

\begin{itemize}
  \item There are still significant differences between OP, OPAL, and more recent opacity tables \citep{2012ApJ...745...10B, 2016ApJ...817..116C, 2016ApJ...821...45K}. Much theoretical work needs to be carried out to diagnose such discrepancies, find any sources of missing opacity, and understand differences with the \citet{2015Natur.517...56B} experiment.   \citet{mor17} calculated two-photon opacity including interference between classes of excited states, angular factors, atomic matrix elements, and density effects. While further investigation is needed before agreement with experiment can be claimed, they obtain substantial cross sections corresponding to an extra opacity comparable with the   \citeauthor{2015Natur.517...56B} measurement. However, \citet{2018HEDP...26...23P} suggests that, due to the intensity of photon sources in the \citeauthor{2015Natur.517...56B} experiment, it is unlikely that two-photon processes would play an important role.

  \item Current computing power and the scale of the physics problem mandate trade-offs between completeness, detail, and accuracy in opacity calculations. We need to include a sufficient number of relevant processes and transitions in sufficient detail to adequately render radiative transfer in stars. The hope has been that, by summing up approximate cross sections over billions of bound--bound and bound--free transitions, we would end up with opacities somewhat more accurate than individual rates. This argument is only valid if all uncertainties are random and uncorrelated, but fails in the case of systematic errors. It is consequently important to identify and assess the importance of any systematic errors in atomic data whether measured or calculated.

  \item Early comparisons of calculated and experimental opacities under astrophysical conditions were made by   \citet{1988ApPhL..52..847D} and \citet{1991PhRvL..67.3784P}; these and subsequent experiments were complemented by measurements of individual cross sections \citep{kjeldsen:Lab-bf-Review}. The experimental opacity determination at the Z-Facility by \citet{2015Natur.517...56B} marked an important advance in so far as conditions close to those at the base of   solar convection zone were attained; the finding of an unexpectedly large iron opacity when compared with all existing theoretical tables has reignited the study of atomic opacities. Experimental reproducibility is of course also a key issue,   and there is a great deal of expectation in independent approaches based on the NIF, (Perry et al., this volume) and on a   double ablation front scheme (Cola\"itis et al., this volume).

  \item Further work is needed in the study of improved EOS for opacities, particularly in the high-density regime. At present the OP, OPAL, and OPLIB opacities use significantly different EOS formulations. This complicates detailed comparisons and the possibility of finding an exact source for the differences among opacity data sets. (See the paper by Kilcrease et al. in this volume.)

  \item In current research on extrasolar planetary atmospheres and planetary formation, forthcoming astronomical observatories, namely, JWST \citep{2006SSRv..123..485G} and ARIEL \citep{2016SPIE.9904E..1XT}, create an enormous need for molecular opacities. Molecules of interest range from simple diatomic systems to organic complexes that can trace the presence of life. These systems yield billions of bound--bound transitions taking significant resources to compute in full detail. Results of these detailed calculations and extensive measurements are available in ever growing databases such as ExoMol \citep{2016JMoSp.327...73T} and HITRAN \citep{jt691s}. (See also papers by Bernath, Calvet, Huang et al., Morley, Tennyson, and Zammit et al. in this volume.)

  \item While a lot of such detailed molecular data are currently available, the complex chemistry of rocky planets with temperatures ${\sim}300\,{\rm K}$ and a potential for life adds thousands of possible molecules for which this   brute-force approach is rendered all but impossible. This has led to the development of alternative methods that are more efficient at producing approximate data (Sousa-Silva et al., this volume).

  \item One of the topics that we had in mind but were not able to cover in the WAO was the computation and measurement of dust opacities. Morley (this volume) shows that the modeling of brown-dwarf and planetary atmospheres to interpret observations must consider the opacity from aerosols (clouds and hazes), for which the optical constants of many compounds (ices, salts, and rocks) are not generally available. The increased wavelength range of the JWST will certainly inflate in the near future the demand for molecular line lists and aerosol opacities. Furthermore, Calvet (this volume) discusses the role dust plays in the structure of protoplanetary disks, as the variations of elemental composition are determined by dust structures (e.g., condensation fronts), and dust opacity depends on grain size distributions that are subject to settling and trapping effects. We would therefore encourage the organizers of the next WAO to promote this important opacity field.
\end{itemize}

%\section{Current needs for astrophysical opacities}

%++++++++++++++++++++++++++++++++++++++++++++++++++++++++++++++++++++++++++++++
\section{How to Move Forward}
%++++++++++++++++++++++++++++++++++++++++++++++++++++++++++++++++++++++++++++++

There are a number of steps in computing opacities for astrophysical entities, and each of these steps needs to be reassessed as part of the refinement of opacity calculations:
\begin{itemize}
  \item[a)] We need detailed, accurate, and complete energy levels, line positions, $f$-values, and cross sections for each process and by each absorber of significance: atoms/ ions/anions/molecules and temporary dipoles formed in collisions. This is the opacity issue that has received the most attention, and great strides have been made as summarized above. In this respect quality assessments and benchmarks of atomic and molecular line data (e.g., the BRASS database project by Lobel et al. and Laverick et al., this volume) by detailed and extensive comparisons of theoretical and observed stellar spectra are essential. Plasma environment effects on close-coupling methods need investigation as a minimum requirement for accuracy to be assured.  See also the {\it Opacity Wish List} (Trampedach, this volume) for new entries needed by the modeling community.

  \item[b)] We expect that the steady increase in computing power will ease the current compromise between accuracy and completeness, allowing both to be exhaustively addressed in future opacity calculations.

  \item[c)] A reliable EOS is needed that includes quantum and relativistic effects, higher order Coulomb interactions (beyond Debye--H{\"u}ckel), realistic microfield distributions and partition functions, and preferably, also molecules beyond H$_2$. Internal consistency results from the physical picture, but this has to be carefully crafted in the chemical picture. The fundamental unsolved problem is the description of atomic states in a plasma environment, which is expected to become increasingly critical at the higher densities.

  \item[d)] The EOS should also provide parameters for line broadening to establish a robust and accurate formulation of line profiles (consistent with the EOS), including Stark and pressure broadening, broadened Fano profiles, and interference by spectator electrons, etc. The importance of line broadening through the Stark effect, as mentioned in Section~3, becomes increasingly important as density increases; line absorption is then redistributed in frequency changing any resulting Rosseland mean opacity depending on them. It is therefore essential to take into account Stark broadening for all relevant L- and M-shell lines. The wavelength range selected must ensure a complete line profile is included in every case.

  \item[e)] We encourage quantum mechanical calculations of the continuum absorption by normal (in the zero-density limit) bound states that are wholly or partially ``dissolved'' by the plasma environment (via the occupation probabilities of the EOS in the chemical picture). Conservation of oscillator strength needs to be ensured.

  \item[f)] Proper accounting of plasma oscillations (through their effect on the dielectric constant) is desirable in the presence of electron degeneracy, screening, and Coulomb interactions.  \citet{2018PhPl...25c2106S}, for example, derive the linear dispersion relation for electromagnetic waves in ultra-relativistic plasmas with arbitrary electron degeneracy that is applicable to white dwarfs and neutron stars.

  \item[g)] Robust and accurate integration of the Rosseland mean including the choice of wavelength or energy scale has often been overlooked. This needs to be done in a way that results in a Rosseland mean that is continuously differentiable in temperature, density, and composition and accompanied by a fast and robust scheme for interpolation in those variables.

  \item[h)] We firmly call for continued helioseismological work to help us constrain the solar chemical composition, the EOS, and atomic radiative data at solar interior conditions. Also further studies of pulsating stars, chemically peculiar stars, and asteroseismology, the latter particularly in the context of hybrid pulsators that have proven to be helpful in elucidating opacity shortcomings.

  \item[i)] As the discrepancies between the measurements by \citet{2015Natur.517...56B} and all current theoretical  opacities remain unexplained, experimentalists are encouraged to continue the analysis of their setups and possible sources of uncertainty \citep{2016PhRvE..93b3202N, 2016HEDP...20...17N, 2017PhRvE..95f3206N}. This compels our community to make a plea for prioritizing experimental work such as that planned at the NIF \citep[][Perry et al., this   volume]{2017JPlPh..83a5903H,2017HEDP...23..223P} and LMJ-PETAL \citep{2015HEDP...17..162L} to determine opacities under a wide range of conditions and independently verify these first experiments, as well as expand their scope.

\end{itemize}

The various indications of problems with current opacity tables as a result of experiments, helioseismology, and asteroseismology surely means the issue impacts a much larger region of stellar structure and evolution, but has hitherto been masked by other uncertainties and not been recognized for what it is. Solving these known opacity problems therefore has the potential to have wide-felt repercussions for astrophysics in general, and of a similar magnitude as the revolution in atomic physics applied to stars as we witnessed around the first WAO.

\acknowledgements
The authors are indebted to Dr. C. A. Iglesias (LLNL) for comments on two drafts of this paper that led to substantial improvements and to Dr. P. Hakel (LANL) for useful suggestions. One of us (AEL-G) is grateful to the University of Oxford for a travel and subsistence allowance that made attendance at the second WAO possible.

%\bibliography{Opac2017}

\end{document}